\documentclass[aps,prl,twocolumn,superscriptaddress,showpacs]{revtex4}

\usepackage{amsmath,bm,amsfonts}
\usepackage{graphicx}

\begin{document}

\title{Combinatorial Search for Optimal Hydrogen-Storage Nanomaterials Based on Polymers} 
\author{Hoonkyung Lee}
\author{Woon Ih Choi}
\author{Jisoon Ihm}
\email[\* corresponding author. Email:\ ]{jihm@snu.ac.kr}
\affiliation{Department of Physics and Astronomy, FPRD, and Center for Theoretical Physics, Seoul National University, Seoul 151-747, Korea}  

\date{\today }

\begin{abstract}
We perform an extensive combinatorial search for optimal nanostructured hydrogen storage materials among various metal-decorated polymers
using first-principles density-functional calculations.
We take into account the zero-point vibration as well as the pressure- and temperature-dependent adsorption-desorption probability 
of hydrogen molecules.
An optimal material we identify is Ti-decorated {\textit {cis}}-polyacetylene with {\textit {reversibly usable}} gravimetric and volumetric 
density of 7.6 weight percent and 63 kg/m$^3$ respectively near {\textit {ambient conditions}}. 
We also propose ``thermodynamically usable hydrogen capacity" as a criterion for comparing different storage materials.
\end{abstract}
 
\pacs{68.43.Bc, 71.15.Nc}

\maketitle
Hydrogen storage is a crucial technology to the development of the hydrogen fuel-cell powered vehicles~\cite{Zuttel,Cratree}.
Recently, nanostructured materials receive special attention because of potentially large storage capacity
(high gravimetric and volumetric density), safety (solid-state storage), and fast filling and delivering
from the fuel tank (short molecular adsorption and desorption time)~\cite{Dillon,Dresselhaus,Rosi}.
However, when the thermodynamic behavior of the gas under realistic environments is taken into account,
the {\textit {usable}} amount of hydrogen with these nanomaterials falls far short of the desired capacity for practical applications
and search for novel storage materials continues worldwide~\cite{Ye,Jhi,Goddard,Serguei}.
It is to be emphasized that hydrogen storage in nanostructured materials utilizes the adsorption  
of hydrogen molecules on the host materials and its thermodynamic analysis
is distinct from that of metal or chemical hydrides. 
Each adsorption site on the nanomaterial behaves more or less independently and the probability of 
the hydrogen adsorption follows the equilibrium statistics which is a smooth function of the pressure and temperature. 
There is no sharp thermodynamic phase transition between the gas and the adsorbed state of H$_{2}$, 
in contrast to the case of metal or chemical hydrides 
where an abrupt phase transition occurs at well-defined pressure at a given temperature~\cite{Chen1}. 

With this caveat, a general formalism applicable to the hydrogen adsorption on nanomaterials was derived in the present study from 
the grand partition function with the chemical potential determined by that of the surrounding H$_{2}$ gas acting as a thermal reservoir. 
As each site can adsorb more than one H$_{2}$ molecule, information on the multiple adsorption energy is necessary. 
(The situation is analogous to the O$_{2}$ adsorption and desorption on hemoglobin which can bind up to 4 O$_{2}$ molecules.) 
In equilibrium of the H$_{2}$ molecules between the adsorbed and desorbed (gas) states, 
the occupation (adsorption) number f is obtained from f=$kT\partial lnZ/\partial \mu$, where $Z$ is the grand partition function, $\mu$ 
is the chemical potential of H$_{2}$ in the gas phase at given pressure {\it p} and temperature {\it T}, and $k$ is the Boltzmann constant.
Here, f per site is reduced to  
\begin{equation}
\mathop{\mathrm{f}} =
\frac{{\sum_{l=0} lg_{l}e^{l(\mu-\varepsilon_{l})/kT}}}{{\sum_{l=0} g_{l}e^{l(\mu-\varepsilon_{l})/kT}}},
\end{equation}
where $\varepsilon_{l}$ is the adsorption energy per H$_{2}$ molecule when the number of adsorbed molecules is $l$ and  
$g_{l}$ is the multiplicity (degeneracy) of the configuration for given $l$.
The summation is over all different configurations up to the maximum number (N$_{\mathrm {max}}$) of adsorbed molecules. 

Another important thermodynamic feature in the H$_{2}$ adsorption energetics is the zero-point vibrations of the H$_{2}$ molecules 
with respect to the host metal atom (e.g. Ti) on which H$_{2}$'s sit. 
(The zero-point vibration {\it within} the H$_{2}$ molecule, on the other hand, exists 
in both gas and adsorbed states and cancels out in the calculation of f.) 
The actual energy $\varepsilon_{l}$ to be used in Eq. (1) is the static adsorption energy (usually calculated and reported in the literature) 
{\it {minus}} the zero-point vibration energy which sums up to as large as 25 $\%$ of the static adsorption energy according to our calculation, 
a value not to be neglected at all.

Considering these thermodynamic aspects, we paid attention to the fact that simple polymers decorated 
with light transition metal atoms may be superior to other recently reported nanomaterials such as Ti-decorated nanotubes~\cite{Yildirim} or 
Sc-decorated fullerenes~\cite{Zhao} in terms of {\it usable} gravimetric and volumetric density. 
The basically one-dimensional nature of polymers is advantageous for compact storage, 
with a very small number of carbon atoms needed to accommodate a decorating metal atom which attracts hydrogen molecules. 
Furthermore, entangled chains of long polymers form a solid structure ideal for safe handling of the hydrogen.
A systematic approach was employed to search for optimized high-capacity hydrogen storage nanostructures based on polymers. 
For the supporting backbone materials, we first considered {\it trans}- and {\it cis}-polyacetylene (among linear carbon chains), 
polyaniline, polyphenol, poly para phenylene, and poly ether ether ketone (chains of hexagonal rings), and 
polypyrrole and polythiophene (chains of pentagonal rings). For decorating transition metals, 
we initially chose all light transition metal elements starting from Sc in the periodic table. 
Various possible adsorption sites of the transition metal atoms were tested for each case. 
The maximum number of adsorbed H$_{2}$ molecules also varied (up to six) at different sites. 
In short, the total combinatorial number in our study exceeded one thousand. 
In practice, we were able to reduce the number considerably by eliminating obviously unfavorable cases using 
a few test calculations of the adsorption energy and structural stability. 
Many kinds of pentagonal and hexagonal ring chains were ruled out. 
For decorating atoms, only Sc, Ti, and V atoms passed the first-round candidate screening test. 
Such a combinatorial search for the optimized material and geometry yielded a few promising nanostructures for hydrogen storage.

We employed spin-polarized first-principles electronic structure calculations based on the density-functional theory~\cite{Kohn}.
The plane-wave based total energy minimization~\cite{Ihm} with the Vanderbilt ultrasoft pseudopotential~\cite{Vanderbilt} was performed.
The generalized gradient approximation (GGA)~\cite{Kim} of Perdew, Burke, and Ernzerhof (PBE)~\cite{PBE} was used in the calculations.
The kinetic energy and the relaxation force cutoff were 35 Ry and 0.001 Ry/a.u., respectively.
For periodic supercell calculations, the distance between polymers was maintained over 10 \AA~in all cases.

The best candidate material we found in our search using the total energy calculations 
was {\it cis}-polyacetylene decorated with Ti atoms whose structure 
after the H$_{2}$ molecule adsorption is presented in Fig. 1(a). 
The binding energy of a Ti atom on this polymer is 2.4 eV. 
The structure has about 2 wt$\%$ higher storage capacity than Ti-decorated {\it trans}-polyacetylene. 
(Since as-synthesized polyacetylene is of {\it cis}-type, it is in principle possible to attach Ti atoms to {\it cis}-polyacetylene although
{\it trans}-polyacetylene is a more stable structure.)
N$_{\mathrm {max}}$ for this structure is five and the H$_{2}$ molecules are compactly adsorbed on both sides of the polyacetylene plane. 
The molecular formula corresponding to this structure is (C$_4$H$_4$$\cdot$2Ti$\cdot$10H$_2$)$_n$.
The maximum gravimetric density (G$_{\mathrm {max}}$) is defined by the weight ratio of 10H$_2$ to C$_4$H$_4$$\cdot$2Ti$\cdot$10H$_2$, 
which is 12 wt$\%$ as shown in Table I. 
G$_{\mathrm {max}}$ for other materials is calculated in the same way. 
We also present other important polymer geometries in Fig. 1 with the maximum number of H$_{2}$ molecules attached to the decorating Ti atoms. 
The calculated (static) adsorption energies per H$_{2}$ as a function of the adsorption number  are presented in Fig. 2 
for easy comparison among different materials. In {\it cis}-polyacetylene, for example, 
they are 0.55, 0.58, 0.48, 0.42, and 0.46 eV/H$_{2}$ for $l$ =1, 2, 3, 4, and 5, respectively. 
As pointed out in previous works, the adsorption of a large number of H$_{2}$ molecules presumably occurs 
through the Dewar-Chatt-Duncanson coordination or Kubas interaction~\cite{Kubas,Niu,Gag}. 
We found the elongation of H$_{2}$ molecules by $\sim$10 $\%$ through electron back donation from 
metal $d$ orbitals to the antibonding hydrogen $s$ orbitals, which supports these theories. 

\begin{figure}[t]
   \centering
   \includegraphics[width=8.6cm]{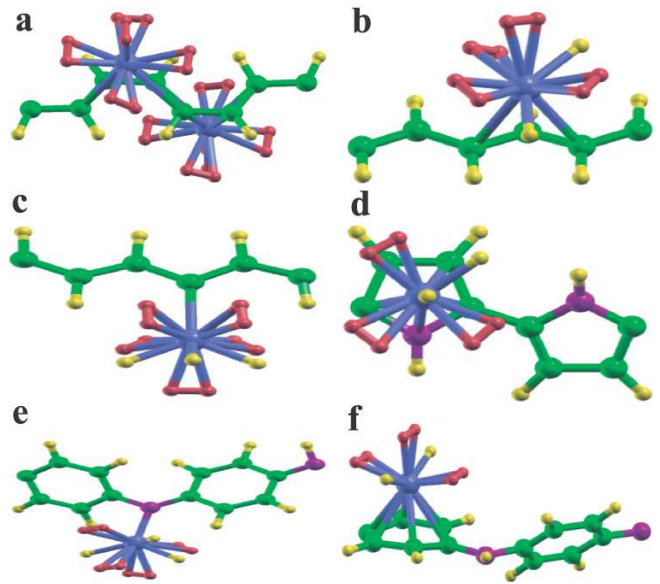}
   \label{Fig.1}
   \caption{ (color online) Atomic structures of the Ti-decorated polymers with the maximum  number of H$_{2}$ molecules attached to Ti atoms. 
    Green, blue, purple, yellow, and red dots indicate the carbon atom, titanium atom, nitrogen atom, hydrogen atom 
    composing the polymer, and the molecular hydrogen, respectively. (a) {\it cis}-polyacetylene with five 
    H$_{2}$ molecules attached per Ti atom. 
    H$_{2}$'s are shown on both sides of the (somewhat distorted) polyacetylene plane. 
    In the rest (b)-(f), H$_{2}$'s are shown only on one side of the polymer for visual clarity. 
    (b) {\it trans}-polyacetylene with Ti atoms located out of the plane of the polymer chain. 
    (c) {\it trans}-polyacetylene with Ti atoms in the plane of the chain. 
    (d) polypyrrole with Ti atoms out of the pentagonal plane. 
    (e) polyaniline with Ti atoms in the hexagonal plane. 
    (f) polyaniline with Ti atoms out of the hexagonal plane.}
\end{figure}
 
 \begin{figure}[t]
    \centering
    \includegraphics[width=8.6cm]{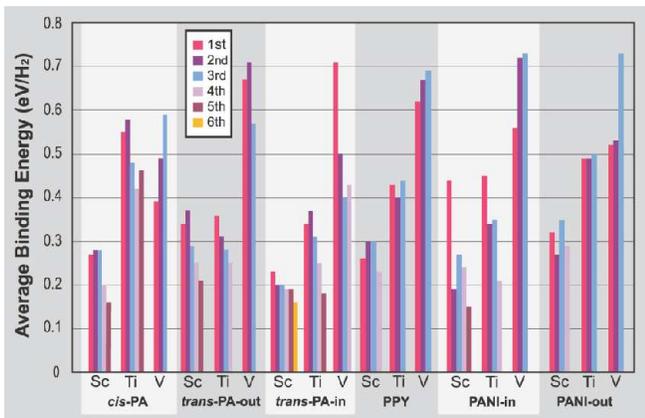}
    \label{Fig.2}
 
    \caption{ (color online) Calculated static adsorption (binding) energy per H$_{2}$ molecule for polymers 
      decorated with Sc, Ti, or V atoms.
      The average binding energy per H$_{2}$ is plotted up to the maximum number of adsorbed H$_{2}$'s allowed for each species. 
      PA, PPY, and PANI stand for polyacetylene, polypyrrole, and polyaniline, respectively. 
      -out and -in mean out-of-plane and in-plane configurations as previously shown in Fig. 1, respectively.}
 \end{figure}

We chose to present in Fig. 2 the {\it static} adsorption energy following the usual practice in the literature~\cite{Zhao,Yildirim}. 
After subtracting zero-point vibration energies (25 $\%$ of the static adsorption energy) for all structures, 
we obtained the true dynamic adsorption energy ($\varepsilon_{l}$) to be used in Eq. (1). 
For instance, the zero-point vibration energy per H$_{2}$ molecule for {\it cis}-polyacetylene was 0.09 eV for H$_{2}$ on top of 
the Ti atom and 0.12 eV for H$_{2}$ attached to the side~\cite{Support}. 
We employed the experimental chemical potential in the literature~\cite{HCP} in the calculation of f. 
The degeneracy factor was approximated by the number of calculated local energy minima for given $l$. 
The largest $g_{l}$ we found was 3 and, since the exponential factor $e^{l(\mu-\varepsilon_{l})/kT}$ dominated, 
$g_{l}$'s turned out to give a minor correction to the result. 
The occupation number f as a function of {\it p} and {\it T} for representative nanomaterials is presented in Fig. 3. 
The occupation-pressure-temperature (f-{\it {p-T}}) diagram of the nanomaterial storage in Fig. 3 is 
the counterpart of the widely-used pressure-composition isotherms (PCI) in metal hydride storage~\cite{Chen1}. 
To obtain the usable amount of hydrogen, it is necessary to specify {\it p} and {\it T} at the time of adsorption (filling) 
and desorption (delivering from the storage tank). 
Since an internationally agreed-upon standard has not been set up, we propose to use the adsorption condition of 30 atm 
and 25 $^{\circ}\mathrm{C}$ and the desorption condition of 2 atm and 100 $^{\circ}\mathrm{C}$, abbreviated to 30-25/2-100. 
These numbers, which may be revised in the future by consensus, are based on information in the literature~\cite{DOE,Bhatia} and 
reflect practical situations in gas filling and vehicles operations. 
30 atm for adsorption and 1.5 atm for desorption were used in Ref. 24, but they did not take advantage of 
the temperature variation. 
\begin{table}[t]
    \begin{ruledtabular}
    \centering
    \caption{\label{tab:double}
    Hydrogen storage capacity of representative nanomaterials from GGA calculations.   
    PA, polyacetylene; PPY, polypyrrole; PANI, polyaniline; CNT, carbon nanotube. 
    All are decorated with Ti except for Sc-decorated C$_{48}$B$_{12}$. -out means an out-of-plane configuration described in Fig. 1. 
    N$_{\mathrm {ads}}$ and N$_{\mathrm {des}}$ are the numbers of attached H$_{2}$'s per site at the condition of adsorption 
    (30 atm-25 $^{\circ}\mathrm{C}$) and
    desorption (2 atm-100 $^{\circ}\mathrm{C}$), respectively. 
    N$_{\mathrm {use}}$ is the practically usable number (N$_\mathrm {{ads}}$$-$N$_{\mathrm {des}}$) 
    and N$_{\mathrm {max}}$ is the maximum number of adsorbed H$_{2}$'s. G and V are gravimetric and volumetric density, respectively.} 
    \begin{tabular}{ccccc}
    {\textbf {Materials}} & N$_{\mathrm {ads}}$-N$_{\mathrm {des}}$ & N$_{\mathrm {use}}$/N$_{\mathrm {max}}$ & 
                            G$_{\mathrm {use}}$/G$_{\mathrm {max}}$ & V$_{\mathrm {use}}$/V$_{\mathrm {max}}$\\  
                                                                       &           &        & {\small (wt$\%$)}   & {\small (kg/m$^3$)} \\
    \hline
    {\textbf {{\textit {cis}}-PA}}                                       & 5.00-1.84 & 3.16/5 & 7.6/12   & 63/100  \\
    {\textbf {PPY}}                                                    & 3.00-0.05 & 2.95/3 & 4.9/5    & 33/34   \\
    {\textbf {PANI-out}}                                               & 3.00-0.96 & 2.04/3 & 4.1/6    & 31/46   \\
    {\textbf {C$_{\bf {48}}$B$_{\bf {12}}$Sc$_{\bf {12}}$}}            & 2.68-0.02 & 2.66/5 & 4.7/8.8  & 23/43   \\
    {\textbf {CNT}}                                                    & 1.95-0.35 & 1.60/3 & 4.1/7.7 &not available\\
  \end{tabular}
   \end{ruledtabular}
\end{table}
\begin{figure}[b]
    \centering
    \includegraphics[width=8.6cm]{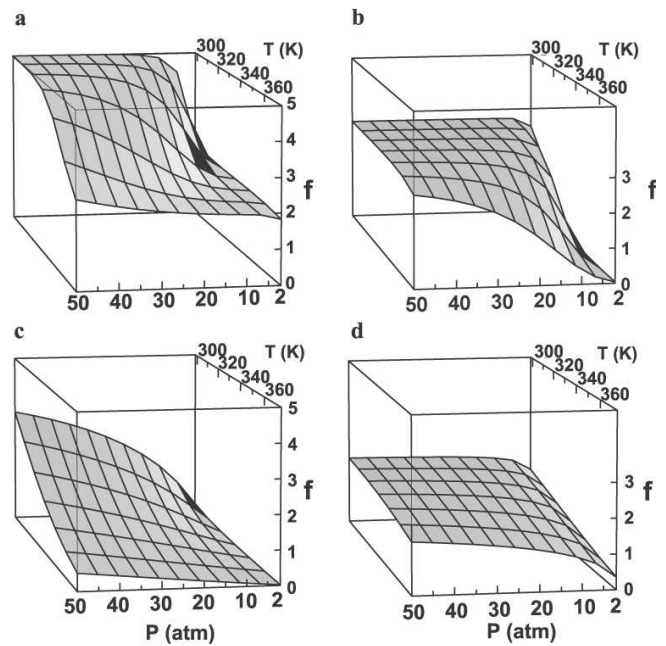}
    \label{Fig.3}
    \caption{Occupation number-pressure-temperature (f-{\textit {p-T}}) diagram of the hydrogen storage in representative nanomaterials. 
      The ranges of the pressure and the temperature cover typical conditions of filling and delivering from the storage tank. 
      (a) Ti-decorated {\it cis}-polyacetylene. 
      (b) Ti-decorated polypyrrole. 
      (c) Sc-decorated C$_{48}$B$_{12}$.  
      (d) Ti-decorated carbon nanotube.}
\end{figure}

We adopted an easily achievable temperature range of 25-100 $^{\circ}\mathrm{C}$ here.
Then, f at the condition of 30-25 minus f at 2-100 is the available number of H$_{2}$ molecules per site. 
These numbers are listed in Table 1. 
For comparison, the same numbers for the Sc-decorated fullerene (C$_{48}$B$_{12}$Sc$_{12}$)~\cite{Zhao} and 
the Ti-decorated carbon nanotube~\cite{Yildirim} are presented as well. 
We confirm that our results for the static adsorption energy of these materials agree with reported values~\cite{Zhao,Yildirim}. 
When converted to the gravimetric density, Ti-decorated {\it cis}-polyacetylene stores {\it usable} H$_{2}$ molecules of 7.6 wt$\%$
out of the {\it maximum} density of 12 wt$\%$, 
which is much greater than, say, the goal of 6 wt$\%$ by the year of 2010 set by the Department of Energy (DOE) of US~\cite{DOE}. 
The Ti-decorated {\it cis}-polyacetylene is the best candidate material for hydrogen storage in the list, 
better than that of Ti-decorated carbon nanotubes~\cite{Yildirim} or Sc-decorated fullerenes~\cite{Zhao}. 
Note that 60 $\%$ desorption of H$_{2}$ is achieved here at a temperature as low as 100 $^{\circ}\mathrm{C}$, 
which is considerably lower than the dissociation temperature of usual metal hydrides. 
If we were to slightly raise the desorption temperature to 130 $^{\circ}\mathrm{C}$, 
the usable gravimetric capacity of {\it cis}-polyacetylene would reach 9 wt$\%$.

The volumetric density of the hydrogen storage is difficult to evaluate and not usually reported in the literature. 
To estimate the volumetric density, we assume that a hydrogen molecule adsorbed at the top of one unit cell and 
another adsorbed at the bottom of the next unit cell are separated by a van der Waals distance ($\sim$3.4 \AA). 
The calculated usable volumetric density is 63 kg/m$^3$, which is higher than the 2010 goal of 45 kg/m$^3$ set by the DOE of US~\cite{DOE}. 

So far, we have demonstrated an enormous potential of polymers as a hydrogen storage medium. 
A serious problem yet to be overcome in practice is the attack of oxygen or other ambient gases. 
Polyacetylene is known to be especially vulnerable to it. 
One possible morphology to avoid oxidation is a dense matrix of polymer which is made permeable to H$_{2}$, but not to O$_{2}$. 
Clustering of decorating metal atoms is another obstacle in material fabrication~\cite{Sun}. 
In our simulation which allowed for the relaxation of atomic positions, individually dispersed metal atoms did show local stability in linear polyaniline. However, how to suppress the aggregation of metal atoms in polymer matrices in general is a difficult question to be answered.
As a brief summary of the experimental situation, we want to point out that  
the hydrogen storage in polymer-dispersed metal hydrides was studied before~\cite{Schmidt}. 
It was hoped that the storage capacity of metal hydrides might be enhanced by incorporating a low-density polymer 
that could interact with the hydride on a molecular level and store additional hydrogen within the polymer structure. 
Although increase in the hydrogen release was found, the maximum amount reached only 0.36 wt$\%$ in experiment. 
We believe that the low capacity is probably due to the poor morphology, i.e., too large ($\geq$10 $\mu m$) Ti-polymer particles. 

In summary, we have carried out first-principles electronic structure calculations for hydrogen binding to metal-decorated polymers of 
many different kinds.
When the thermodynamic behavior of hydrogen molecules under realistic conditions is considered, the Ti-decorated {\it cis}-polyacetylene
is found to have the highest {\it usable} gravimetric and volumetric density among nanostructures reported so far.
We also propose the f-$p$-$T$ diagram as a criterion for evaluating usable capacity at ambient conditions.
It remains to be a challenge for experimentalists to fabricate a structure of individually dispersed Ti atoms on polymer
as much as possible in order to achieve significantly improved storage capacity. 

We acknowledge the support of the SRC program (Center for Nanotubes and Nanostructured Composites) of 
MOST/KOSEF and the Korea Research Foundation Grant No. KRF-2005-070-C00041. Computations are performed through 
the support of KISTI.

\end{document}